\documentclass[aps,superscriptaddress,showpacs,twocolumn]{revtex4}

\usepackage[dvips]{graphicx}
\usepackage{amssymb}

\def \beq {\begin{equation}}
\def \eeq {\end{equation}}

\begin{document}
\title{Lyapunov statistics and mixing rates for intermittent systems}

\author{Carlos J.A. Pires}\email{carlosamado@ig.com.br}
\address{Centro de Matem\'atica, Computa\c c\~ao e Cogni\c c\~ao, UFABC, 09210-170, Santo Andr\'e, SP, Brazil}
\author{Alberto Saa}\email{asaa@ime.unicamp.br}
\address{Departamento de Matem\'atica Aplicada, UNICAMP, 13083-859, Campinas, SP, Brazil}
\author{Roberto Venegeroles}\email{roberto.venegeroles@ufabc.edu.br}
\address{Centro de Matem\'atica, Computa\c c\~ao e Cogni\c c\~ao, UFABC, 09210-170, Santo Andr\'e, SP, Brazil}

\date{\today}

\begin{abstract}
We consider here a recent conjecture stating that correlation functions and tail probabilities of finite time Lyapunov exponents would have the same power law decay in weakly chaotic systems. We demonstrate that this conjecture fails for a generic class of maps of the Pomeau-Manneville type. We show further that, typically,
the decay properties of
 such tail probabilities  do not provide significant information on key aspects of weakly chaotic dynamics  such as ergodicity and instability regimes. Our approaches are firmly based on rigorous results, particularly the Aaronson-Darling-Kac theorem, and are also confirmed by exhaustive numerical simulations.
\end{abstract}

\pacs{05.45.Ac, 02.50.Ga, 74.40.De}

\maketitle

\section{Introduction}
\label{sec1}

It has been well known since the seminal works of Sinai, Ruelle, and Bowen (SRB) \cite{Sinai,Bowen,Ruelle} that the strongest chaotic systems (Smale's axiom A and Anosov systems) have SRB measures  with exponentially decaying mixing rates (see also Ref. \cite{SRB}). For these systems,
 the difference
  between temporal and spatial averages is statistically
  described by
   a Gaussian distribution (the central limit theorem) and   the convergence
   of both averages towards a unique value is assured.
On the other hand, there is a wide range of systems where mixing rates
and other related correlation functions
decay as   power laws. Such class of dynamical systems,     dubbed { weakly chaotic} in the Physics literature, typically
exhibits weak statistical properties when compared with the chaotic ones.
Examples of   weakly chaotic systems include maps with indifferent fixed points \cite{indf,Sarig,Gouzel,PS}, billiards \cite{bil}, and Hamiltonian systems with sticky islands in phase space \cite{hst}, among  others.
These systems have  in common an intermittent dynamical
behavior, exhibiting a transition  from regular to chaotic regimes which
has attracted the attention of physicists and mathematicians in the last 20 years. We recall that
the so-called mixing rate of a pair of  phase space  observable
functions $\phi$ and $\psi$
for maps of the type
\beq
\label{map}
x_{t+1}=f(x_{t})
\eeq is
defined as being the correlation function
\begin{eqnarray}
\label{mix}
C_t(\phi,\psi) &=& \left| \int \phi(x)\psi\left(f^t(x)\right)\, d\mu(x)
\right. \\
&& \left. - \int   \phi(x)\, d\mu(x) \int  \psi(x)\, d\mu(x) \right|,\nonumber
\end{eqnarray}
where $\mu(x)$ is the invariant measure under the map $f$.
A map   is said to be { mixing} if $C_{t}\to 0$ as $t\to\infty$ for
any pair of  phase space smooth observables  $(\phi,\psi)$.

Since correlation functions  as the mixing rate (\ref{mix}) might characterize the transition from  strong to weak chaos, the attempt to relate them to more fundamental dynamical quantities is a goal that looks, at first glance, really promising. In Ref. {\cite{ALP}}, for instance, the finite time Lyapunov exponents
\beq
\label{lyap}
\Lambda_{t}(x)=\frac{1}{t}\sum_{k=0}^{t-1}\ln\left|f'\left(f^{k}(x)\right)\right|,
\eeq
were considered for one-dimensional maps like (\ref{map}) for which $\Lambda_{\infty}>0$, {\em i.e.}, for strongly chaotic cases. Essentially, they show that if there exist two positive constants $\Lambda_0$ and $\gamma>1$ such that
\begin{eqnarray}
\label{distlamb}
M_{t}(\Lambda_{0})=\int_{\Lambda_{0}}^{\infty}\eta(\Lambda_{t})d\Lambda_{t}\le a_{1}t^{-\gamma},
\end{eqnarray}
where $\eta(\Lambda_{t})$ stands for the distribution of finite time Lyapunov exponents for the system in question,
then we also have the following upper bound for the mixing rates
\begin{eqnarray}
\label{corupp}
C_{t}(\phi,\psi)\le a_{2}t^{-(\gamma-1)},
\end{eqnarray}
of any pair of H\"older continuous observables $(\phi,\psi)$. In other words, they have proved rigorously that if the finite time Lyapunov exponents tail probability is bounded  by $t^{-\gamma}$ for large $t$, then correlations will be also bounded asymptotically by $t^{-(\gamma-1)}$.
These important results have inspired a recent work \cite{AM} in which a related conjecture is made for weakly chaotic systems, {\em i.e.}, irrespective of having $\Lambda_{\infty}>0$.  The  main results of \cite{AM} can be summarized as follows.

\begin{enumerate}
\item They argue that scrutinizing the way in which $1-M_{t}$ decays for large $t$ provides an ``extremely efficient way'' of studying quantitatively the decay of correlations.
\item They conjecture, in view of recent results of \cite{ALP}, that the estimates of (\ref{distlamb}) and (\ref{corupp}) are not optimal, and the decay properties of $1-M_{t}$ and of $C_{t}-\int\phi d\mu\int\psi d\mu$ should be, in fact, both polynomial with the same exponent.
\item They check numerically such a conjecture for an one-dimensional intermittent map with two indifferent fixed points of the Pomeau-Manneville type \cite{PM}, for which the polynomial decay rates of correlations are known exactly. Two other types of two-dimensional intermittent maps are also numerically considered to support the conjecture.
\end{enumerate}

Here, we show that this conjecture is false by presenting an explicit class of counter-examples. We will
consider a general class of Pomeau-Manneville maps \cite{PM} and show that their Lyapunov exponents tail probability (\ref{distlamb}) decay faster than any power law, whereas  their correlations (\ref{mix}) do
exhibit a power law  decay. Our approaches are firmly based on rigorous results, particularly on the Aaronson-Darling-Kac theorem \cite{Aaronson}, and are also confirmed by exhaustive numerical simulations. For all maps in this class, correlation functions decay slowly than the Lyapunov exponents tail probability, suggesting  that  bounds of the type (\ref{distlamb}) and (\ref{corupp}) can be  physically relevant
also for  weakly chaotic systems.

\section{Lyapunov Statistics}
\label{sec2}

Our counter-examples consist in a general class of Pomeau-Manneville (PM) intermittent dynamical systems of type (\ref{map}) with $f:[0,1]\rightarrow[0,1]$, where
\beq
\label{pmap}
f(x)\sim x(1+ax^{z-1})
\eeq
for $x \to 0$, with $a>0$ and $z>1$. The global form of $f$ is irrelevant, provided it respects the axioms of an AFN-system \cite{RZ}. For maps of the type (\ref{pmap}), $x=0$ is an indifferent (neutral) fixed point, {\em i.e.}, $f(0)=0$ and $f'(0)=1$. Such systems are known to have power law invariant measures near their indifferent fixed points. More specifically, we have $d\mu(x)=\omega(x)\,dx$, where $\omega(x)\sim bx^{-1/\alpha}$ near the fixed point $x=0$, with $\alpha=(z-1)^{-1}$ \cite{Thaler}. As consequence, such systems have diverging invariant measure near this points for $z>2$. Moreover, finite invariant measure ($1<z<2$) implies ergodicity and the usual Lyapunov exponential instability, whereas the diverging case ($z>2$) implies nonergodicity and subexponential instability. We will consider each of these cases separately and   show that the conjecture proposed in \cite{AM} fails for both.

\subsection{Exponential instability}

Let us first consider the statistics of finite time Lyapunov exponents (\ref{lyap}) for randomly distributed initial conditions $x\in [0,1]$, in the case of finite invariant measure cases ($1<z<2$). It is well known that ergodicity properties can determine completely
such statistics. For instance,  Birkhoff theorem \cite{GDB} states that,
in an ergodic regime,
the time average of an arbitrary observable function $\vartheta$,   $t^{-1}\sum_{k=0}^{t-1}\vartheta(f^{k}(x))$, converges uniformly to the spatial average $\int\vartheta\,d\mu$. Then, for almost all initial conditions $x\in[0,1]$, the local expansion rate $\Lambda_{t}(x)$ in (\ref{lyap}) converges to the unique positive Lyapunov exponent $\Lambda_{\infty}$ as $t\rightarrow\infty$. On the other hand, if $t$ is finite, $\Lambda_{t}(x)$ assumes different values depending on the initial condition $x$. The corresponding probability density function $\eta(\Lambda_{t})=\eta(\Lambda,t)$ is given by
\beq
\label{plamb}
\eta(\Lambda,t)=\int\delta(\Lambda_{t}(x)-\Lambda)d\mu(x).
\eeq
For large $t$, $\eta(\Lambda,t)$ takes the scaling form \cite{Ellis}
\beq
\label{pesc}
\eta(\Lambda,t)\sim \eta(\Lambda_{\infty},t)\exp[-t\Omega(\Lambda)],
\eeq
where $\Omega(\Lambda)\geq0$ is a concave function with minimum at $\Omega(\Lambda_{\infty})=0$. Then we have $\Omega(\Lambda)\sim c_{1}(\Lambda-\Lambda_{\infty})^{2}$ and $\eta(\Lambda_{\infty},t)\sim(c_{1}t/\pi)^{1/2}$, with $c_{1}>0$. Now, a simple calculation  by using Laplace's method leads to
\beq
\label{Mt0}
M_{t}\sim\frac{1}{2}\mbox{erfc}(\sqrt{\Omega_{0}t}),
\eeq
where $\Omega_{0}=\Omega(\Lambda_{0})$. The decaying properties of Eq. (\ref{Mt0}) are definitively different than those ones predict by \cite{AM}. In fact, one has
\beq
\label{Mt1}
M_{t}\sim\frac{1}{2\sqrt{\pi}}\frac{\exp(-\Omega_{0}t)}{\sqrt{\Omega_{0}t}},
\eeq
for $\Lambda_{0}\neq\Lambda_{\infty}$ and $M_{t}(\Lambda_{\infty})\sim 1/2$. It is important to stress that there are many rigorous results in the literature establishing polynomial bounds for the decay of correlations of Pomeau-Manneville maps in the regime $1<z<2$, see \cite{indf} and references therein. Most notably, Sarig \cite{Sarig} and Gou\"ezel \cite{Gouzel}  have achieved optimal polynomial bounds for such correlations in this regime. Therefore, contrary to the conjecture proposed in \cite{AM}, polynomial decay of correlations can occur simultaneously  with exponential decay of Lyapunov tail probability distributions. In fact, any conjecture stating that $M_{t}$ should decay as a power law is generically violated for ergodic regimes. For the PM map with $1<z<2$, this is indeed predicted by
Theorem 2 (exponential level I result) of \cite{PS}.

\subsection{Subexponential instability}

Let us consider now the cases for which the invariant measure diverges locally at the indifferent fixed point $x=0$, {\em i.e.}, $z>2$. For such cases, the system typically exhibits a nonergodic behavior and, hence, time averages do not converge to a unique constant value. Nevertheless, the Aaronson-Darling-Kac (ADK) theorem \cite{Aaronson,JA,DK} ensures that a suitable time-weighted average does converge uniformly in distribution terms towards a Mittag-Leffler
distribution of unit first moment. More specifically, for a positive function $\vartheta$ and a random variable $x$ with
an  absolutely continuous measure with respect to the Lebesgue measure
on the interval $[0,1]$, there is a (return) sequence $\left\{a_{t}\right\}$ for which
\beq
\frac{1}{a_{t}}\sum_{k=0}^{t-1}\vartheta(f^{k}(x))\stackrel{d}{\longrightarrow}\xi_{\alpha}\int\vartheta d\mu
\label{DAC}
\eeq
for $t\to\infty$, where $\xi_{\alpha}$ is a non-negative Mittag-Leffler random variable of index $\alpha\in(0,1)$ and with unit expected value.
The return sequence $\left\{a_{t}\right\}$ for PM systems like (\ref{pmap})
and  $0<\alpha<1$ is given by \cite{RZ,TZ}
\beq
\label{at}
a_{t}\sim\frac{1}{b}\frac{1}{a}\left(\frac{a}{\alpha}\right)^{\alpha}\frac{\sin(\pi\alpha)}{\pi\alpha}t^{\alpha},
\eeq
for $t\rightarrow\infty$. What the ADK theorem is really pointing out here is to the explicit necessity of dealing  with finite time subexpoential Lyapunov exponents
\beq
\lambda_{t}^{(\alpha)}(x)=\frac{1}{t^{\alpha}}\sum_{k=0}^{t-1}\ln\left|f'\left(f^{k}(x)\right)\right|
\label{lamba}
\eeq
instead of the usual  exponents (\ref{lyap})   for PM systems of the AFN type    (see also Ref. \cite{AAKB}). From Eq. (\ref{DAC}), we have
\begin{eqnarray}
\frac{\lambda_{t}^{(\alpha)}}{\left\langle \lambda\right\rangle}\stackrel{d}{\longrightarrow}\xi_{\alpha},
\label{lml}
\end{eqnarray}
for $t\rightarrow\infty$, where the ADK average value $\left\langle \lambda\right\rangle$ is given by
\begin{eqnarray}
\label{Lac}
\left\langle \lambda\right\rangle=\frac{1}{ba}\left(\frac{a}{\alpha}\right)^{\alpha}\frac{\sin(\pi\alpha)}{\pi\alpha}\int_{0}^{1}\ln|f'(x)|\omega(x)\,dx.
\end{eqnarray}

The ADK theorem  completely determines the correlations and the
tail probability of
 Lyapunov exponents
for the maps of
the type (\ref{pmap}), as one  can   see  by considering a randomly
distributed initial condition $x\in[0,1]$ with probability density $h(x)>0$
in Eq. (\ref{DAC}), leading to
\beq
\label{eq1}
\frac{1}{a_t}\sum_{k=0}^{t-1}\int \vartheta\left(f^k(x)\right) h(x)\, dx =
\int\vartheta\, d\mu
\eeq
for $t\to\infty$. We can rearrange this expression and write
\beq
C_t(\phi,\vartheta) -\int\phi\,d\mu \int\vartheta\,d\mu 
\sim  \alpha \langle\vartheta \rangle t^{\alpha-1} ,
\label{corr}
\eeq
for $t\to\infty$,
where $\phi(x)= h(x)/\omega(x)$ and the  ADK average $\langle\vartheta \rangle$
is   given by an  expression analogous to Eq. (\ref{Lac}). 
 It remains now to show that
the tail probability of
 Lyapunov exponents  for systems
of the type (\ref{pmap}) does not decay as predicted by Eq. (\ref{corr}).
From Eq. (\ref{lml}),
one can obtain $M_t$  for the
map (\ref{pmap})
 by recalling that $\lambda_{t}^{(\alpha)}=t^{1-\alpha}\Lambda_{t}$, implying the
following
distribution of finite time Lyapunov exponents for systems of
 type (\ref{pmap})
\beq
\label{eta}
\eta\left(\Lambda_t \right) = \frac{t^{1-\alpha}}{\langle\lambda\rangle}\rho^{(r)}_\alpha\left(
\frac{t^{1-\alpha}\Lambda_t}{\langle\lambda\rangle} \right),
\eeq
where $\rho^{(r)}_\alpha$ is a Mittag-Leffler probability density function with unit first moment, which corresponds to choice $r^\alpha = \alpha \Gamma( \alpha)$, according to the definitions of \cite{SV}.
Then, we have finally from Eqs. (\ref{distlamb}) and (\ref{eta})
\beq
\label{mti}
M_{t}=\int_{u(t)}^{\infty}\rho^{(r)}_\alpha(s)\, ds,
\eeq
for $t\to\infty$, where $u(t) = t^{1-\alpha}\Lambda_0/\left\langle \lambda\right\rangle$. The behavior of $\rho^{(r)}_\alpha(x)$ for large
$x$ was recently discussed in \cite{SV}, based on the known
relation between
Mittag-Leffler and one-sided L\'evy distributions \cite{Feller} and
the Mikusinski's asymptotic analysis \cite{Mik} of the latter. In particular,
one has
\beq
\label{asympML2}
\rho_\alpha^{(r)}(x) \sim
\sqrt{\frac{A}{2\pi\alpha}}
  \frac{x^{(2\alpha-1)/(2-2\alpha)}}{1-\alpha} {\exp{\left(-Ax^{1/(1-\alpha)}\right)}},
\eeq
for $r^\alpha = \alpha\Gamma( \alpha)$,
valid for  $x\to\infty$, where
 \beq
 A = \frac{1-\alpha}{\alpha}\Gamma(\alpha)^{1/(\alpha-1)}.
 \eeq
 The integral of Eq. (\ref{asympML2}) can be written in terms of the complementary error function, leading simply to
\beq
\label{decay1}
M_{t}\sim \frac{1}{\sqrt{2\alpha}}\, \mbox{erfc}\left(
\sqrt{Bt}
\right),
\eeq
for large $t$, where
\beq
B= \frac{1-\alpha}{\alpha}\left(
\frac{\Lambda_0\Gamma(\alpha)}{\langle\lambda\rangle}\right)^{1/(\alpha-1)}.
\eeq

The decaying properties of Eq. (\ref{decay1}) are also definitively different than those ones predict by Eq. (\ref{corr}), in the context of conjecture proposed in \cite{AM}. Once more we have
\beq
\label{cauda}
M_{t}\sim
 \frac{1}{\sqrt{2\pi\alpha}}\frac{\exp (-Bt)}{\sqrt{B t }},
\eeq
for large $t$,
demonstrating finally that the conjecture presented in \cite{AM} is false. It is noteworthy that Lyapunov exponents tail probability given by (\ref{cauda}) is essentially the same we would expect from finite measure cases, {\em i.e.}, Eq. (\ref{Mt1}). This shows that the way in which $M_{t}$ decays does not provide significant information on key aspects of weakly chaotic dynamics.

\section{Numerical simulations}
\label{sec3}

In order to test and illustrate
the conclusions of the last Section, we perform an exhaustive numerical analysis of two particular AFN-maps of the type (\ref{pmap}), namely the Thaler map \cite{Thaler}, defined for $z> 2$ as
\begin{eqnarray}
f(x)=x\left[1+\left(\frac{x}{1+x}\right)^{z-2}-x^{z-2}\right]^{-1/(z-2)},
\label{tmap}
\end{eqnarray}
mod 1, and the   modified Bernoulli map  \cite{AAKB}, defined also for
$z>2$ as
\beq
\label{bern}
f(x) = \left\{
\begin{array}{ll}
\displaystyle x + 2^{z-1}x^z,  & 0 \le x \le \displaystyle \frac{1}{2}, \\
\displaystyle x - 2^{z-1}(1-x)^z, & \displaystyle \frac{1}{2} < x \le 1.
\end{array}
\right.
\eeq
The Thaler map (\ref{tmap}) is very convenient here because
its  invariant measure density is explicitly known, namely \cite{Thaler}
\begin{eqnarray}
{\omega}(x)=x^{-1/\alpha}+(1+x)^{-1/\alpha},
\label{rhowt}
\end{eqnarray}
where $\alpha=(z-1)^{-1}$, allowing in this way the explicit evaluation of the ADK averages like (\ref{Lac}). In contrast with the Thaler map, there is no explicit expression for the invariant measure of the modified Bernoulli map, but it is known to have the form $\omega(x)\sim b_{k}|x-x_{k}|^{-1/\alpha}$, also with $\alpha = (z-1)^{-1}$, in the neighborhood of each of the two indifferent fixed points $x_{0}=0$ and $x_{1}=1$ \cite{Tinv}. Note that the ADK theorem is also valid for systems with more than one indifferent fixed point \cite{RZ}. For the Bernoulli map (\ref{bern}), we also have return rates in the form $a_{t}\sim t^{\alpha}$ for $0<\alpha<1$ \cite{RZ}. However, since its corresponding explicit expression for the invariant measure is lacking, we cannot evaluate the ADK averages for the modified Bernoulli map. We will show that this problem can be circumvented by exploiting the numerical data.

\subsection{Tail probability of Lyapunov exponents}

\begin{figure}[tb]
\includegraphics[width=1\linewidth]{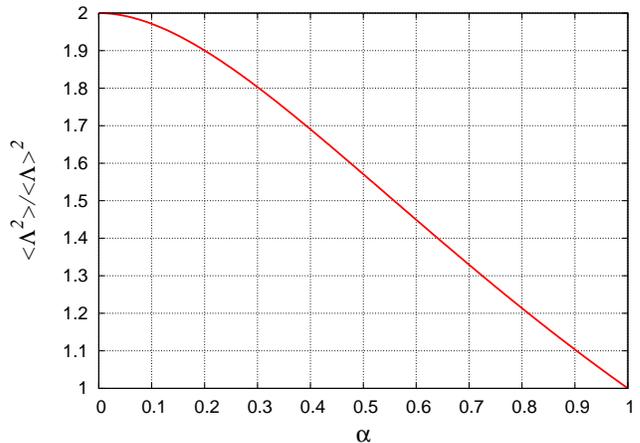}
\caption{Graphics of $\langle \Lambda^2  \rangle/\langle \Lambda  \rangle^2$
as a function of $\alpha$ for Mittag-Leffler
distributions with unit first moment, see Eq. (\ref{constr2}). The Mittag-Leffler index $\alpha$
can be determined from the value of
$\langle \Lambda^2  \rangle/\langle \Lambda  \rangle^2\in [1,2]$.}
\label{fig0}
\end{figure}

Our first task here is to determine if the tail probability of finite time Lyapunov exponents (\ref{distlamb}) for the maps (\ref{tmap}) and (\ref{bern}) do indeed decay
as predicted by Eq. (\ref{decay1}). From Sect. II, we known that such decaying behavior
is assured if the the distribution of finite time Lyapunov exponents were
effectively
given by Eq. (\ref{eta}). We compute numerically the distribution of finite time Lyapunov exponents $\Lambda_t$ for the maps (\ref{tmap}) and (\ref{bern}) for random
initial condition and large $t$, and confront the obtained numerical
data with the
theoretically predicted distribution (\ref{eta}).
The algorithm for the numerical computation of Mittag-Leffler distributions
with arbitrary index $\alpha$ introduced in \cite{SV} was instrumental
to perform such task. The key point of our analysis is to check if a given distribution is well described or not by a generic  Mittag-Leffler probability density function.
We recall   that  a
Mittag-Leffler probability density
$\rho^{(r)}_\alpha(x)$  is defined from its Laplace transform as
\beq
\label{laplace1}
\int_0^\infty e^{-sx}\rho^{(r)}_\alpha(x)\, dx=
\sum_{n=0}^{\infty}\frac{(-sr^{\alpha})^{n}}{\Gamma(1+n\alpha)},
\eeq
for $s\ge 0$, with $0<\alpha< 1$. The choice $r^\alpha = \alpha\Gamma(\alpha)$ assures that $\langle x\rangle = 1$, where the
average here is evaluated with respect to $\rho^{(r)}_\alpha(x)$. From Eqs. (\ref{eta}) and (\ref{laplace1}), we have the following constraints on the high order moments
\beq
\label{constr}
\frac{\langle \Lambda^n  \rangle}{\langle \Lambda  \rangle^n} =
\frac{n!\alpha^{n-1}\Gamma(\alpha)^{n}}{n\Gamma(n\alpha) }
\eeq
of the probability density (\ref{eta}).
One can evaluate $\langle \Lambda^n \rangle$ easily from the numerical
data and
the constraints (\ref{constr}) can be objectively used to decide if
a given distribution is well described or not by a Mittag-Leffler
probability density. In particular, notice that one can determine the
two free parameters of the distribution (\ref{eta}),
 $\langle\lambda \rangle$
and $\alpha$, by considering, for instance, $\langle \Lambda  \rangle =
t^{\alpha-1}\langle \lambda  \rangle$
and
\beq
\label{constr2}
\frac{\langle \Lambda^2  \rangle}{\langle \Lambda  \rangle^2} =
\frac{ \alpha \Gamma(\alpha)^2}{ \Gamma(2\alpha) }.
\eeq
It is very instructive to inspect the graphics of
$\langle \Lambda^2  \rangle/\langle \Lambda  \rangle^2$ as a
function of $\alpha$, see Fig. \ref{fig0}.
For  Mittag-Leffler distributions with unit first moment, one has necessarily
$1 \le \langle \Lambda^2  \rangle/\langle \Lambda  \rangle^2 \le 2$, with
the boundaries corresponding, respectively, to $\alpha=1$ and $\alpha=0$.
For such values of $\alpha$, the Mittag-Leffler probability
density  function
approaches, respectively, a $\delta$-function centered in $x=1$ and a
simple exponential $e^{-x}$, see \cite{SV}. The violation of such boundaries would point out
  unequivocally that one is not leading with  Mittag-Leffler
distributions with first unit moment.
Analogous bounds hold also for higher order moments (\ref{constr}),
$1 \le \langle \Lambda^n  \rangle/\langle \Lambda  \rangle^n \le n!$.

For the case of the Thaler map,  both parameters  $\langle\lambda \rangle$
and $\alpha$ in the distribution (\ref{eta}) are predicted theoretically
by the ADK theorem, allowing the inspection of the convergence rate of Eq. (\ref{lml}) with respect to $t$ and to the number of initial conditions
used to evaluate the Lyapunov exponents.  On the other hand, for the
modified Bernoulli map one cannot determine  exactly  the average
$\langle\lambda \rangle$, but it is possible to infer its value by
\begin{figure}[tb]
\includegraphics[width=1\linewidth]{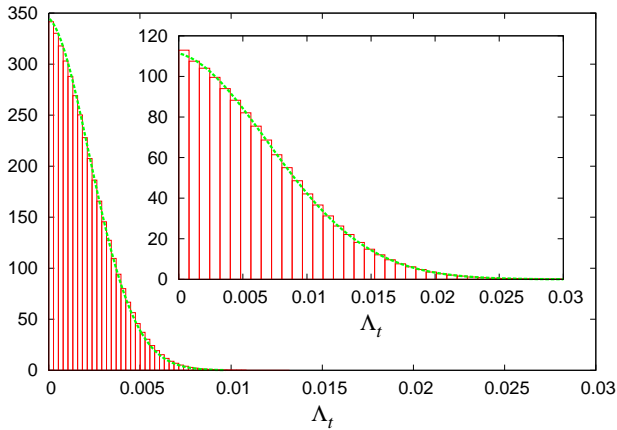}
\caption{Distribution of finite time Lyapunov exponents
 (\ref{lyap})
 for the Thaler map,
determined from the iteration of Eq. (\ref{tmap}), with $z=22/7$
($\alpha = 7/15$), for $2.5\times 10^5$ initial conditions
uniformly distributed on the interval $[0,1]$. The histograms are
built directly from the numerical data, while
 the solid lines are the corresponding Mittag-Leffler
 probability density (\ref{eta}), computed by means of
  the algorithm of Ref. \cite{SV}.
The inset and the background plots correspond, respectively,
 to    $t=6\times 10^4$ and $t=5\times 10^5$.}
\label{fig1}
\end{figure}
computing
$\langle\Lambda \rangle$ from the numerical data and then using $\left\langle \lambda\right\rangle=t^{1-\alpha}\left\langle \Lambda\right\rangle$.
Fig. \ref{fig1}
\begin{figure}[tb]
\includegraphics[width=1\linewidth]{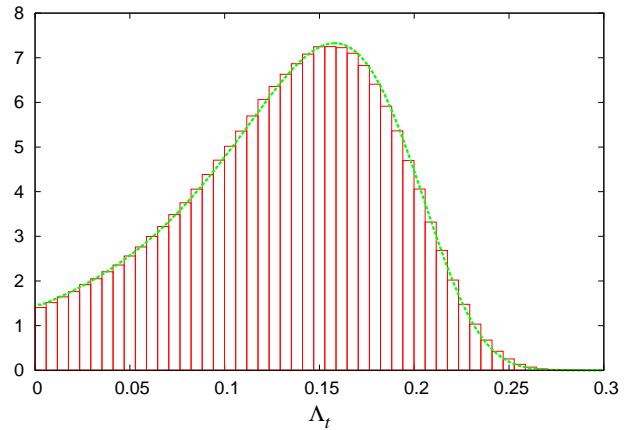}
\caption{Distribution of finite time Lyapunov exponents  (\ref{lyap}) for the modified
Bernoulli map, determined from the iteration of Eq. (\ref{bern}),
with $z=28/13$ ($\alpha=13/15$) and $t=6\times 10^4$, for $2.5\times 10^5$
initial conditions uniformly distributed on the interval $[0,1]$.
The histogram was built
directly from the numerical data, the solid line is  the
 Mittag-Leffler probability density computed
 with the algorithm of Ref. \cite{SV}. The behavior of the distribution
for large $t$ is identical to the Thaler map case depicted in Fig. \ref{fig2}.
In particular,  one also has $\langle \Lambda^n \rangle\to 0$ for $t\to\infty$ and $\Lambda_\infty = 0$, in agreement with the predictions of Eq. (\ref{eta}).}
\label{fig2}
\end{figure}
depicts the
distribution of finite time Lyapunov exponents for the Thaler map (\ref{tmap}).
The plots show clearly that the distribution of
 Lyapunov exponents becomes  peaked around the
origin for $t\to\infty$, leading to $\langle \Lambda^n \rangle\to 0$
for large $t$. In particular, one has $\Lambda_\infty = 0$, in perfect
agreement with the
predicted distribution (\ref{eta}) and the
fact that the Thaler map is known to be weakly
chaotic.  Fig. \ref{fig2}
illustrates the case of the modified Bernoulli map
 (\ref{bern}). For all cases, we see, graphically and according to the
 higher order moments constraints (\ref{constr}), that the distribution of
 finite time Lyapunov exponents is very well described by a Mittag-Leffler probability
 density according to the prediction of Eq. (\ref{eta}). The tail probability
 (\ref{decay1}) is then guaranteed  for these maps.

\begin{figure}[tb]
\includegraphics[width=1\linewidth]{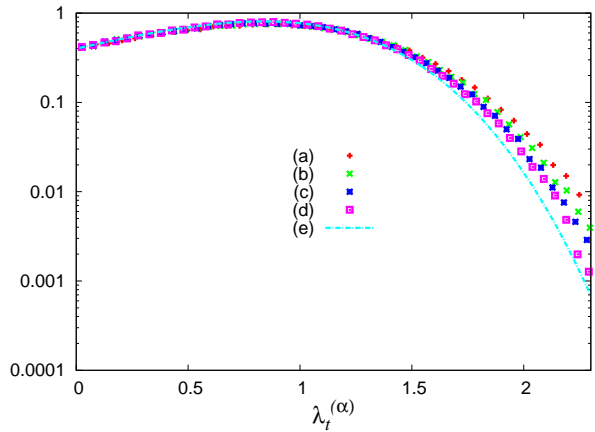}
\caption{Log-plot of the
distribution of finite time subexponential
 Lyapunov exponents (\ref{lamba})  for Thaler map,
determined from the iteration of Eq. (\ref{tmap}),
with $z=26/11$ ($\alpha=11/15$),
for $2.5\times 10^5$ initial conditions uniformly
distributed on the interval $[0,1]$.
The set of points (a), (b), (c), and (d) correspond, respectively, to
the histograms built
 from the numerical data obtained for $t=10^4$, $t=5\times 10^4$,
 $t=25\times 10^4$, and $t=10^6$.
   Each one of these data sets are very well described by a
 Mittag-Leffler probability
 density, see Table \ref{table}.
 The line (e)
  corresponds to the Mittag-Leffler probability
 density
 with the ADK   values for $\alpha$ and
 $\langle\lambda\rangle$.
As one can see, the numerically obtained distributions converges towards
the prediction of the ADK theorem with the increasing of $t$. The increasing
of the number of initial condition does not alter considerably such
convergence, but a better description of the density tail requires
a larger
number of initial conditions, as expected.
}
\label{fig3}
\end{figure}

Our numerical examples are, in fact, illustrating the convergence of Eq. (\ref{lml}), which is a consequence of ADK theorem. As expected, for large
values of $t$ and for large numbers of initial conditions, the histograms
of both Figs. \ref{fig1}  and \ref{fig2}
approach the Mittag-Leffler probability density with the
theoretical predicted values of $\alpha$ and $\langle \lambda\rangle$.
We could, however, detect another very interesting
property. For a given value of $t$ and a given number of initial conditions,
the
corresponding histograms are already very well described by
 a Mittag-Leffler probability density!
With the increasing of $t$ and the number of initial condition,
such ``instantaneous'' Mittag-Leffler probability density approaches
the ADK ones, as it is illustrated, for the Thaler map, in Fig. \ref{fig3}
and in Table \ref{table}.
\begin{table}[t]
\begin{center}
\begin{tabular}{|c|c|c|c|c|c|}
\hline
\hline
 & (a) & (b) & (c)  & (d) & ADK \\
\hline
\hline
$\langle \lambda   \rangle $    & 0.853 & 0.840 & 0.829 & 0.822 & 0.807\\
\hline
$\langle \lambda^2  \rangle/\langle \lambda  \rangle^2$   & 1.305 & 1.303 & 1.300& 1.296 & 1.290\\
\hline
$\langle \lambda^3  \rangle/\langle \lambda  \rangle^3$   &  1.956 & 1.947 & 1.935& 1.921 & 1.899\\
\hline
$\langle \lambda^4  \rangle/\langle \lambda  \rangle^4$   & 3.216 & 3.189 & 3.155& 3.117 & 3.051\\
\hline
$\langle \lambda^5  \rangle/\langle \lambda  \rangle^5$   & 5.674 & 5.598 & 5.510& 5.414 & 5.242\\
\hline
$\langle \lambda^6  \rangle/\langle \lambda  \rangle^6$   & 10.589 & 10.383 & 10.168& 9.927 & 9.497\\
\hline
$15\alpha$   & 10.81 & 10.84 & 10.88 & 10.92 & 11 \\
\hline
\hline
\end{tabular}
\caption{\label{table} Statistical data for the graphics in Fig. \ref{fig3}.
For each data set, the higher order moments constraints (\ref{constr})
are respected, showing that each ``instantaneous'' histogram of Fig. \ref{fig3}
is indeed well described by a Mittag-Leffler probability density. The values
of the Mittag-Leffler index $\alpha$ for each one of the data sets
(last row)
were calculated
from $\langle \lambda^2  \rangle/\langle \lambda  \rangle^2$, see Eq. (\ref{constr2}) and also Fig. \ref{fig0}. The last column corresponds
to the probability density predicted from the ADK theorem, namely the
curve (e) in Fig. \ref{fig3}.}
\end{center}
\vspace{-0.6cm}
\end{table}
The solid lines in both Figs. \ref{fig1}  and \ref{fig2}, for instance,
are the instantaneous
Mittag-Leffler probability densities, {\em i.e.}, their parameters
$\alpha$ and $\langle\lambda \rangle$, although close to the theoretically
predicted values, were calculated from the numerical data by using $\left\langle \lambda\right\rangle=t^{1-\alpha}\left\langle \Lambda\right\rangle$ and Eq. (\ref{constr2}). Table \ref{table} shows the
values of the higher order moments constraints (\ref{constr}) for the
data sets presented in Fig. \ref{fig3}.

\subsection{Correlation functions}

\begin{figure}[tb]
\includegraphics[width=1\linewidth]{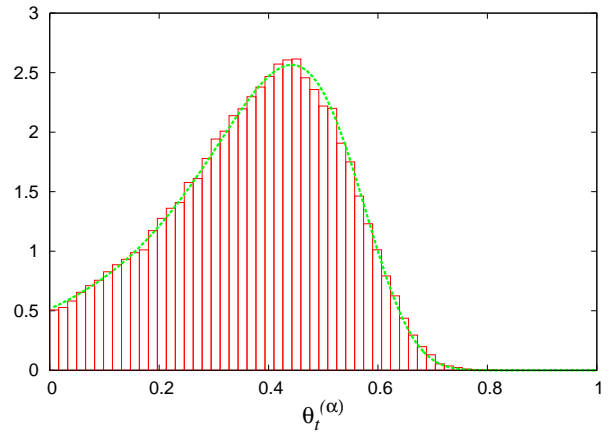}
\caption{Distribution of the quantity $\theta_t^{(\alpha)}$, given by Eq. (\ref{theta_t}), for $\vartheta(x)=\sin^z \pi x$,
determined from the iteration of modified Bernoulli map (\ref{bern}),
with $z=28/13$ ($\alpha=13/15$) and $t=6\times 10^4$, for $2.5\times 10^5$
initial conditions uniformly distributed on the interval $[0,1]$.
The histogram is built
directly from the numerical data and
the solid line  is the corresponding
 Mittag-Leffler probability density, computed by means of
   the numerical algorithm of Ref. \cite{SV}.
}
\label{fig4}
\end{figure}

The numerical computation of the correlation functions   (\ref{mix}) is
rather tricky for the maps in question
 due to the highly discontinuous nature of the
iterated maps $f^t(x)$ for large $t$.
For both cases (\ref{tmap}) and (\ref{bern}), for instance, the iterated map
$f^t(x)$ has $2^{t}-1$ discontinuous points. An accurate numerical
computation   for large $t$  of a correlation function as $C_t$ would
require an extremely fine subdivision of the interval $[0,1]$, rendering
the task  practically and computationally   unviable.
 Nevertheless,
in order to establish the correlation function decaying  (\ref{corr}), it is enough to assure that, for some value of $0<\alpha<1$, the quantity
\beq
\label{theta_t}
\theta_t^{(\alpha)} = \frac{1}{t^{\alpha}}\sum_{k=0}^{t-1} \vartheta\left(
f^k(x)
\right),
\eeq
where $\vartheta(x)$ is
  an integrable function,
converges uniformly in distribution terms towards a random variable for
large
$t$. The ADK theorem does assure such a convergence with
$\alpha=(z-1)^{-1}$ for the maps in question, and, as
demonstrated in Section 2, the correlation decaying (\ref{corr}) is also firmly
based on the ADK theorem. As an example, let us consider the observable function
$\vartheta(x)=\sin^z \pi x$ for the case of the modified Bernoulli map.
According to the discussion of Section II, the correlation $C_t(h,\vartheta)$ for any smooth observable function $h(x)>0$ will exhibit
 a power lay decay as predicted
by Eq. (\ref{corr}) provided $\theta_t^{(\alpha)}$ does converge in distribution terms to a Mittag-Leffler
random variable.
Fig. \ref{fig4} depicts
such distribution and one can confirm the
very good agreement with the predictions of the ADK theorem, assuring the
validity of the correlation decaying (\ref{corr}).
 Similar results hold also for the Thaler map (\ref{tmap}) and
for other observables.

We notice that the validity of Eq. (\ref{corr}) is stronger than the ADK theorem, in the sense that the convergence to a Mittag-Leffler distribution
given by Eq. (\ref{DAC}) is not a necessary condition to establish
 Eq. (\ref{eq1}). In fact, the existence of a  sequence $a_t\sim t^{\alpha}$ such
that ${a_{t}^{-1}}\sum_{k=0}^{t-1}\vartheta(f^{k}(x))$ does converge
in distribution terms towards a random variable, not necessarily of the
Mittag-Leffler type, is enough to assure the decaying (\ref{corr}).

\section{Final remarks}
\label{sec4}

We close by noticing  that the first map presented  in \cite{AM} to
support  the conjecture we have just proved to be false is also a
map with indifferent fixed points,
 namely the so-called Pikovsky map, which is
defined implicitly by \cite{Pik}
\beq
\label{pik}
x = \left\{
\begin{array}{ll}
\displaystyle\frac{1}{2z}\left(1 + f(x) \right)^z, & 0 < x < \displaystyle\frac{1}{2z}, \\
\displaystyle f(x) + \frac{1}{2z}\left(1 - f(x) \right)^z, & \displaystyle\frac{1}{2z} < x < 1.
\end{array}
\right.
\eeq
 The Pikovsky map is defined on the interval $[-1,1]$. For negative $x$, one has simply $f(x) = - f(-x)$. This map
 has two indifferent
fixed points located at $x=\pm 1$ for $z>1$. The correlation functions
 for the Pikovsky map
are known to decay as a power law \cite{Pik,PikMR}. The authors of \cite{AM}
present some numerical evidence suggesting that the tail probability
$M_t$ for the map (\ref{pik}) would also decay  with the same
 power law. This fact seems
to contradict our results of Section II. However, a closer inspection of Eq. (\ref{pik}) reveals that the Pikovsky map is not an AFN-map \cite{RZ} and, hence, the ADK theorem cannot be invoked here
to determine the distribution of
finite time Lyapunov exponents. From the first equation of map (\ref{pik}),
we have
\beq
\frac{f''}{\left(f'\right)^2} = (1-z)\left(2zx \right)^{-1/z},
\eeq
showing that the  axiom A (Adler's condition) \cite{RZ} is not satisfied for $x=0$ and positive $z$. The violation of Adler's condition here is related
to the infinity slope of the map at the origin, and this is known to be
capable of inducing some new dynamical properties as, for instance, the
existence of a regular invariant measure in spite of the
indifferent fixed points,  see Example 1 of \cite{Z1}, for instance.
The failure of Adler's condition  might explain why the authors of \cite{AM} have arrived to the conclusion that
$M_t$ decays as a power law  for the map (\ref{pik}), but certainly a
deeper investigation of the Pikvosky map  would be
interesting and revealing.

\acknowledgments

This work was partially
supported by CNPq (AS and RV), FAPESP (AS), and UFABC (CJAP).
RV wishes to thank V. Pinheiro for enlightening discussions.

\end{document}